# Spatially Distributed Soil Moisture From Traveltime Observations of Crosshole Ground Penetrating Radar using Markov chain Monte Carlo Simulation

Niklas Linde[1], and Jasper A. Vrugt[2,3]

[1]Institute of Geophysics, University of Lausanne, 1015 Lausanne, Switzerland.

[2]Department of Civil and Environmental Engineering, University of California Irvine, 4130 Engineering Gateway, Irvine, CA 92697-2175, USA.

[3]Institute for Biodiversity and Ecosystems Dynamics, University of Amsterdam, Amsterdam, The Netherlands.

**Abstract**

Geophysical methods offer several key advantages over conventional subsurface measurement approaches, yet their use for hydrologic interpretation is often problematic. Here, we introduce theory and concepts of a novel Bayesian approach for high-resolution soil moisture estimation using traveltime observations from crosshole Ground Penetrating Radar (GPR) experiments. The recently developed Multi-try DiffeRential Evolution Adaptive Metropolis with sampling from past states, MT-DREAM$_{(ZS)}$ is being used to infer, as closely and consistently as possible, the posterior distribution of spatially distributed vadose zone soil moisture and/or porosity under saturated conditions. Two differing and opposing model parameterization schemes are being considered, one involving a classical uniform grid discretization and the other based on a discrete cosine transformation (DCT). We illustrate our approach using two different case studies involving geophysical data from a synthetic water tracer infiltration study and a real-world field study under saturated conditions. Our results demonstrate that the DCT parameterization yields the most accurate estimates of distributed soil moisture for a large range of spatial resolutions, and superior MCMC convergence rates. In addition, DCT is admirably suited to investigate and quantify the effects of model truncation errors on the MT-DREAM$_{(ZS)}$ inversion results. For the field example, lateral anisotropy needs to be enforced to derive reliable soil moisture variability. Our results also demonstrate that the posterior soil moisture uncertainty derived with the proposed Bayesian procedure is significantly larger than its counterpart estimated from classical smoothness-constrained deterministic inversions.

**Introduction**

Spatial model regularization is required to ensure numerical tractability, uniqueness and stability of finely-discretized deterministic geophysical inverse problems. In fact, these properties inspired Occam's inversion (e.g., Constable et al., 1987) that seeks to find the most parsimonious model structure that fits the measurements of the system under consideration as closely and consistently as possible with remaining error residuals that accurately mimic the underlying noise characteristics of the data. By construction, such models have a spatially-variable resolution that is significantly coarser than the discretization of the inverse model grid, and depend on the type of model regularization being used. The importance of this resolution discrepancy has been largely over-looked in past publications —but also in more recent contributions to the hydrogeophysics literature that use geophysical tomograms for hydrologic calibration or property characterization (e.g., Rubin et al., 1992; Hubbard et al., 1999; Chen et al., 2001; Farmani et al., 2008). These resolution characteristics limit the direct use of deterministic geophysical tomograms for hydrologic model construction and predictions as theoretical or laboratory-derived petrophysical relationships are not applicable to the typically overly smooth tomographic models (e.g., Day-Lewis and Lane, 2004; Day-Lewis et al., 2005; Moysey et al., 2005; Linde et al., 2006b). In fact, for non-linear inverse problems and for applications to vadose zone or fractured rock systems with large and sharp variations in physical properties and state variables, it is virtually impossible to assess the reasonableness and accuracy of the derived models. Furthermore, the linearized parameter uncertainty intervals derived from Occam's inversions are strongly dependent on the regularization used to create a stable solution (Alumbaugh and Newman, 2000). This leads to overly optimistic estimates of model uncertainty, particularly in cases when model smoothness is imposed based on arguments of mathematical convenience rather than solid prior information. These problems can be partly resolved by simultaneous (joint) use of multiple types of geophysical data (e.g., Linde et al., 2008) or by using additional and/or other information contained in the geophysical signal (e.g., Ernst et al., 2007).

A promising and alternative parameter estimation approach that avoids the need to directly relate geophysical tomographic images to hydrologic properties is to solve a hydrologic inverse problem instead by direct coupling of the geophysical data to hydrologic model parameters and state variables using suitable petrophysical relationships and forward modeling (e.g., Kowalsky et al., 2005; Hinnell et al., 2010). However, this approach supposes availability of an adequate conceptual model of the subsurface. Furthermore, the hydrologic boundary conditions are assumed to be largely known *a-priori*, although it is possible to extend the inversion to jointly estimate these forcing variables as well (e.g., Kowalsky et al., 2005). Most applications of hydrogeophysics and environmental geophysics focus on hydrologic characterization of the subsurface when quantities such as rainfall, evaporation, and soil water fluxes are largely unknown. In all such applications it is fundamental to know how well the geophysical data themselves constrain the hydrologic properties and state variables of interest at a given spatial resolution before they can be used for hydrologic purposes. This makes it necessary to invert the geophysical data.

A variety of approaches have been proposed in the literature to investigate the variance and resolution properties of deterministic inverse problems. One pragmatic and rather popular approach is to simply perform several independent inversions with different regularization terms and reference models. This provides a qualitative assessment of how well the parameters are resolved by the available geophysical data (Oldenburg and Li, 1999). An alternative approach is to use most-squares inversion (Jackson, 1976; Meju and Hutton, 1992), in which the tolerable ranges of a given model parameter are sought for by allowing small increases in the data misfit. Another important contribution constitutes the work of Kalscheuer and Pedersen (2007) who investigated the relationship between the uncertainty of a given parameter and its corresponding spatial resolution. Although these different methods provide information about model resolution and parameter uncertainty, they provide limited insight into the underlying probability distribution of the model parameters and their multi-dimensional cross-correlation. This information is of utmost importance, particularly if, for example, the soil moisture estimates derived from geophysical tomograms are to be used as "calibration data" in a hydrologic inversion (e.g., Farmani et al., 2008).

In this paper, we present a novel Bayesian method devoid of these limitations and reverse the original question of Kalscheuer and Pedersen (2007), to back out at a given uniform spatial resolution the corresponding variability of each model parameter. The resulting model variability and spatial covariance can then, at a later stage, be used for hydrologic modeling and inversion purposes. The method that we propose is general in the sense that it could be applied with slight modifications to most geophysical techniques that are commonly used in vadose zone and groundwater studies. This involves, amongst others, electrical resistance tomography, crosshole seismics, surface refraction seismics, frequency- and time-domain electromagnetics, and induced polarization. For simplicity, we focus on soil moisture estimation from travel time measurements of crosshole ground penetrating radar (GPR). Crosshole GPR has become a popular measurement technique in vadose zone hydrology to obtain in situ estimates of soil moisture and its spatial distribution (e.g., Eppstein and Dougherty, 1998; Binley et al., 2001; Alumbaugh et al., 2002; Kowalsky et al., 2005). The first-arrival traveltime is arguably easiest to retrieve yet contains imminent information about the spatial distribution of the soil moisture throughout the vadose zone. Specifically, each traveltime refers to the time-lapse between the emission of an electromagnetic pulse from a transmitter antenna to the first-arriving energy at a receiver antenna. This traveltime is strongly related to (1) soil moisture along the travel path of the electromagnetic energy, and (2) the path length that depends on the geometry of the experiment and the in situ soil moisture distribution. In this paper, we illustrate our methodology by application to two-dimensional (2-D) soil moisture estimation. Extension to three-dimensions (3-D) is straightforward, but computationally more challenging.

Most applications of crosshole GPR data in the vadose zone use asymptotic solutions of the wave equation governing electromagnetic (EM) radiation (see Klotzsche et al. (2010) for an exception using full-waveform inversion). The EM energy is then modeled as traveling along an infinitely thin curved ray between two points (i.e., the centers of the transmitter and receiver antennas). Ray-bending in the presence of soil heterogeneity is often ignored (e.g., Binley et al., 2001; Kowalsky et al., 2005; Hansen et al., 2008) so that a linear relation between the geophysical data and model suffices, but with the risk of obtaining biased model estimates.

Previous studies have employed stochastic concepts to interpret crosshole GPR data. For example, Chen et al. (2001) used a linear regression model derived from collocated borehole data and geophysical tomographic estimates to update kriged hydraulic conductivity fields using Bayesian theory. Hansen et al. (2006) used sequential Gaussian simulation and linear theory to derive geostatistical realizations honoring a given geostatistical model, as well as borehole and geophysical data. Another study by Gloaguen et al. (2007) used cokriging and cosimulation to create multiple radar slowness models for a linearized theory, in which the model covariance matrix was also inversely estimated. To create geostatistical realizations consistent with large-scale features resolved by deterministic inversion, Dafflon et al. (2009) used perturbations conditioned on GPR tomograms to obtain porosity models from simulated annealing that honor borehole data and an assumed geostatistical model.

In this paper, we introduce a Bayesian framework for spatially distributed soil moisture estimation using traveltime observations from crosshole GPR. This methodology has several distinct advantages over prevailing deterministic and stochastic inversions. While classical deterministic inversions only provide an estimate of the most likely soil moisture values, our method combines recent advances in adaptive Markov chain Monte Carlo sampling (Vrugt et al., 2008, 2009; Laloy and Vrugt, 2012), with an Eikonal solver to estimate the posterior probability density function (pdf) of spatially distributed soil moisture. Our method (1) requires limited information about the initial soil moisture distribution; (2) directly incorporates the non-linear relation between models and simulated data; and (3) explicitly treats individual error sources, including those associated with the petrophysical model and traveltime data. The posterior soil moisture distribution derived with our framework provides a means to derive (spatially variable) soil hydraulic properties, and estimates of vadose zone model parameter, state, and prediction uncertainty. Moreover, the size of the posterior soil moisture uncertainty constitutes an important diagnostic measure to help judge the information content of the first-arrival traveltime data for hydrologic inversion and analysis. Computational limitations impose a ray-based approach, which discards useful information contained in the acquired GPR traces and might lead to modeling errors that bias the inversion results. With continued advances in computational resources, these assumptions can be relaxed by using numerical models that accurately resolve antenna radiation patterns and EM wave propagation (e.g., Lambot et al., 2004; Ernst et al., 2006; Warren and Giannopolous, 2011).

This work is directly related to the recent contribution of Laloy et al. (2012) who investigated whether 3-D soil moisture plumes can be accurately constrained by crosshole GPR traveltimes using a model parameterization based on Legendre moments. This previous study assumed explicit prior knowledge of the morphological features of the 3-D target moisture distribution. The present 2-D study is more concerned with an analysis of the trade-off between model variance and resolution for cases without prior information about the soil moisture distribution. We consider two different and opposing model parameterization schemes: one based on uniform grids with constant cell properties and one based on a spectral representation using the discrete cosine transform.

To test our methodology, we first apply the method to a synthetic infiltration study in a heterogeneous soil in which the 2-D soil moisture distribution is exactly known (Kowalsky et al., 2005). This is followed by a real-world study using geophysical field data from the South Oyster Bacterial Transport Site in Virginia (Hubbard et al., 2001; Linde et al., 2008; Scheibe et al., 2011). Throughout this paper, we compare the soil moisture uncertainty estimates derived with our Bayesian methodology against those from a classical deterministic Occam's inversion and first-order uncertainty analysis.

**Theory**

*Definition of the Stochastic Inverse Problem*

In this paper, we estimate spatially distributed maps of soil moisture from crosshole GPR traveltime observations. Each map constitutes a random draw from the posterior samples derived with MCMC simulation using MT-DREAM$_{(zs)}$ (Laloy and Vrugt, 2012). This set of samples is created by constructing multiple different Markov chains in parallel that generate a random walk through the search space and successively visit solutions with a stable frequency stemming from the target distribution of interest. Particularly, our framework derives the posterior pdf of soil moisture $\boldsymbol{\theta}$, $p(\boldsymbol{\theta}|\mathbf{t})$ given measurements of crosshole GPR traveltimes, hereafter referred to as $\mathbf{t}$. For didactic and computational purposes, we explicitly differentiate between two successive steps that involve (1) geophysical inversion and (2) petrophysical transformation. The first, and arguably most important step in view of the present paper, derives the posterior distribution, $p(\mathbf{m}|\mathbf{t})$ of the geophysical property of interest, $\mathbf{m}$ using Bayes law

$$p(\mathbf{m}|\mathbf{t}) = \frac{p(\mathbf{m})p(\mathbf{t}|\mathbf{m})}{p(\mathbf{t})}, \qquad [1]$$

where $p(\mathbf{m})$ signifies the prior distribution, and $L(\mathbf{m}|\mathbf{t}) \equiv p(\mathbf{t}|\mathbf{m})$ denotes the likelihood function that summarizes the statistical properties of the error residuals in a single scalar value. The normalization constant, or evidence, $p(\mathbf{t})$ is difficult to estimate directly in practice, and is instead derived from numerical integration

$$p(\mathbf{t}) = \int p(\mathbf{t}|\mathbf{m})p(\mathbf{m})d\mathbf{m} \qquad [2]$$

over the feasible space $\mathbf{M}$; $\mathbf{m} \in \mathbf{M} \in \mathbb{R}^M$ so that $p(\mathbf{m}|\mathbf{t})$ scales to unity. Explicit knowledge of $p(\mathbf{t})$ is desired for Bayesian model selection and averaging, but otherwise all statistical inferences (mean, standard deviation, etc.) of $\mathbf{m}$ can be made from the unnormalized density

$$p(\mathbf{m}|\mathbf{t}) = p(\mathbf{m})L(\mathbf{m}|\mathbf{t}). \qquad [3]$$

The user is free to select any functional shape of $p(\mathbf{m})$, but in the absence of detailed prior information, a uniform hypercube (also called flat, noninformative or uniform prior) is typically used. Many geophysical parameters, such as radar velocity are so-called Jeffrey parameters, which means that their values are strictly positive and that their reciprocal property might as well be used as inversion parameter (slowness in the case of velocity; resistivity in the case of conductivity, etc.). A uniform distribution of a Jeffrey parameter provides different results than a uniform distribution between the same upper and lower bounds using its reciprocal property. To make the parameterization independent of the two physically equivalent formulations (e.g., velocity and slowness) we use a logarithmic transformation (e.g., Tarantola, 2005). Consequently, $\mathbf{m}$ refers to a logarithmic transformation of the geophysical properties of interest.

To estimate, the posterior moisture distribution, $p(\mathbf{m}|\mathbf{t})$ in Eq. [3] an explicit definition of $L(\mathbf{m}|\mathbf{t})$ is required to judge the "distance" of each model simulation to the actual measurement data. To simplify the analysis somewhat we make the common assumption that the measurement errors of $\mathbf{t}$ are homoscedastic (constant variance), approximately Gaussian, and spatially uncorrelated. This leads to the following formulation of the log-likelihood $l(\mathbf{m}|\mathbf{t})$

$$l(\mathbf{m}|\mathbf{t}) = -\frac{n}{2}\ln(2\pi) - \frac{n}{2}\ln(\sigma^2) - \frac{1}{2}\sigma^{-2}\sum_{i=1}^{n}(g_i(\mathbf{m}) - t_i)^2, \qquad [4]$$

where $n$ denotes the total number of geophysical observations, $\sigma$ signifies the measurement error, and $\mathbf{g}(\mathbf{m})$ represents the prediction (simulation) of the forward model. The first and second term at the right-hand side are constants whose values do not depend on the actual parameter values, **m**. The third term is the weighted sum of squared error, traditionally being used in many model calibration studies. In the absence of detailed information about the variance of the error, $\sigma^2$, we estimate this value along with the model parameters. This is a common approach in statistics, and results presented in Bikowski et al. (2012) demonstrate the usefulness of this methodology to estimate the measurement error of dispersive GPR data. Note that we only invert for a single composite measurement error that summarizes the effects of all sources of uncertainty. It would be possible to disentangle $\sigma$ into its constituent error sources, but this requires some prior information about the probabilistic properties of each individual error term. The assumption of spatially uncorrelated errors is likely violated in the presence of geometrical and modeling errors. This can be resolved by using a different formulation of the likelihood function, but is outside the scope of the present study. The first-arrival traveltimes are computed with the finite-difference (FD) algorithm *time3d* (Podvin and Lecomte, 1991). The solution is obtained in the high-frequency limit by considering a point source and a first-order approximation of the Eikonal equation.

Once $p(\mathbf{m}|\mathbf{t})$ is known, a petrophysical model is used to derive the posterior soil moisture distribution, $p(\theta|\mathbf{t})$ using

$p(\theta|\mathbf{t}) = L(\theta|\mathbf{m})p(\mathbf{m}|\mathbf{t}),$ [5]

where $L(\theta|\mathbf{m})$ signifies the likelihood of observing $\theta$ given **m**. This two step approach given by Eqs. [3] and [5] differentiates explicitly between errors that originate from the geophysical measurements, inherent non-uniqueness of the model dimension, experimental design and underlying physics that are summarized in $p(\mathbf{m}|\mathbf{t})$ and errors arising from the petrophysical relationship that are described in $L(\theta|\mathbf{m})$.

The functional form of $L(\theta|\mathbf{m})$ depends on the geophysical method being used, the geological setting, available data, and the scale at which the relationship is to be employed. A significant body of literature exists that has investigated the (petrophysical) relationship between electrical permittivity (the underlying physical property that determines the radar velocity) and soil moisture (e.g., Topp et al., 1980; Roth et al., 1990). The choice of $L(\theta|\mathbf{m})$ should accurately reflect the actual field and small-scale variability of the petrophysical relationships. For real-world applications, this choice can be a challenging task. In this paper, we consider the moisture uncertainty to be determined by variations and measurement errors of radar velocity only; additional uncertainty can be expressed by allowing the petrophysical relationships to vary as function of space/scale using additional calibration parameters in $L(\theta|\mathbf{m})$.

*Markov Chain Monte Carlo with MT-DREAM$_{(zs)}$*

To generate samples from the posterior target distribution, $p(\mathbf{m}|\mathbf{t})$ we use MCMC simulation with the MT-DREAM$_{(zs)}$ algorithm. This method runs multiple chains simultaneously in parallel and uses multi-try proposal sampling from an archive of past states to explore the posterior target distribution. A detailed description of the MT-DREAM$_{(ZS)}$ algorithm appears in *Laloy and Vrugt* [2012], and interested readers are referred to this publication. Jumps in each chain $i = 1,…,N$ are generated by taking a fixed multiple of the difference of two or more randomly chosen members (chains) of an archive of $M$ past states, $\mathbf{Z}$ (without replacement) [*Vrugt et al.*, 2008, 2009]:

$$\mathbf{m}_{new}^{i} = \mathbf{m}^{i} + \left(\mathbf{1}_{d} + \mathbf{e}_{d}\right)\gamma\left(\delta,d'\right)\left[\sum_{j=1}^{\delta}\mathbf{Z}^{r_1(j)} - \sum_{h=1}^{\delta}\mathbf{Z}^{r_2(h)}\right] + \boldsymbol{\varepsilon}_{d}, \qquad [6]$$

where $\delta$ signifies the number of pairs used to generate the proposal, and $r_1(j)$, $r_2(h) \in \{1,…,M\}$; $r_1(j) \neq r_2(h)$ for $j = 1, …, \delta$ and $h = 1, …, \delta$. The values of $\mathbf{e}_d$ and $\boldsymbol{\varepsilon}_d$ are drawn from $U_d(-b,b)$ and $N_d(0,b^*)$ with $b$ and $b^*$ small compared to the width of the target distribution, respectively, and the value of jump-size $\gamma$ depends on $\delta$ and $d'$, the number of dimensions that will be updated jointly. The Metropolis acceptance probability is used to decide whether to accept candidate points or not

$$\alpha = \min\left(1,\exp\left(l\left(\mathbf{m}_{new}|\mathbf{t}\right) - l\left(\mathbf{m}|\mathbf{t}\right)\right)\right). \qquad [7]$$

If the proposal is accepted the chain moves to $\mathbf{m}_{new}$ otherwise the chain remains at its current (old) position. Following the recommendations of *Laloy and Vrugt* (2012) we use $N = 3$ with five parallel proposals in each chain. Furthermore, we use $\gamma = k_\gamma / \sqrt{(2d)}$, with $k_\gamma = 0.5$. With a probability of 20% we temporarily set $\gamma = 1.0$ to allow direct jumps between disconnected posterior modes (ter Braak, 2006).

The MT-DREAM$_{(zs)}$ approach solves two important problems in MCMC sampling. First, the algorithm automatically selects an appropriate scale and orientation of the proposal distribution en route to the target distribution. Second, heavy-tailed and multimodal target distributions are efficiently accommodated, as MT-DREAM$_{(zs)}$ directly uses the past locations of the chains, instead of their covariance, to generate candidate points, allowing the possibility of direct jumps between different modes. The algorithm is particularly designed for distributed computing, and receives high sampling efficiencies compared to other MCMC algorithms (Laloy and Vrugt, 2012). The convergence of the joint chains is diagnosed with the Gelman-Rubin convergence statistic (Gelman and Rubin, 1992).

*Model Parameterization and Prior Ranges*

The parameterization of the inverse model should ideally provide enough details to capture all relevant variations in the Earth's property field and thereby avoid bias caused by model truncation (Trampert and Snieder, 1996). In global earth seismology, it is widely recognized that differences between deterministic inversion results based on the same data set are largely due to differences in spatial or spectral resolutions used in the (inverse) parameterizations (e.g., Chiao and Kuo, 2001). Compactly supported pixels that in a Cartesian coordinate system would be represented by, for example, uniform grid models provide a very high spatial resolution, whereas spherical harmonics or discrete cosine transform (DCT) in its Cartesian counterpart provide a high spectral resolution. Many popular model parameterization schemes can be found in-between these two opposing extremes, such as wavelets (e.g., Chiao and Kuo, 2001) and slepian functions (e.g., Simons et al., 2006). In the following, we compare uniform grid and DCT parameterizations.

The uniform grid parameterization has the key advantage of a localized and uniform spatial resolution. Inversion based on a uniform grid distribution is hence straightforward. One simply defines a uniform grid size of the model parameters (the logarithm of slowness) using lower and upper bounds $m_{min}$ and $m_{max}$ for each individual parameter $m_i$, $i = \{1,\ldots,d\}$. Key disadvantages of this approach include (1) problems in forward simulation due to the presence of discrete boundaries between property values of neighboring cells (e.g., Podvin and Lecomte, 1991); (2) a strong sensitivity to small variations of the model boundaries; (3) visually suspect inversion results.

Model regularization and deterministic inversion based on the DCT (Ahmed et al., 1974) was recently introduced in the hydrologic and geophysical literature (Jafarpour and McLaughlin, 2008; Jafarpour et al., 2009) and resolves many of the problems associated with uniform grid parameterizations. The DCT has a number of advantages for stochastic inversion, for instance (1) the resolution and separation of scales is explicitly defined, (2) the transformation is orthogonal and close to the optimal Karhunen-Loève transform, (3) the computational efficiency is high, (4) the basis vectors depend only on the dimensionality of the model, and (5) the transform is linear and operates with real parameter values. The 1-D

DCT-II (hereafter referred to as DCT) of a uniformly discretized model **x** with discretization length $\Delta x$ is

$$G(k) = \beta(k) \sum_{p=0}^{P-1} x(p) \cos\left(\frac{\pi(2p+1)k}{2P}\right), \qquad [8]$$

where

$$\beta(k) = \begin{cases} \dfrac{1}{\sqrt{P}} & k = 0 \\ \sqrt{\dfrac{2}{P}} & 1 \leq k \leq P-1 \end{cases}, \qquad [9]$$

where $G(k)$ are the DCT coefficients, and $P$ is the size of the model. The first basis defines the constant background, whereas subsequent bases describe variations around this value at increasingly higher frequencies. For example, Figure 1 in Jafarpour et al. (2008) depicts the results for the first 8 × 8 DCT bases and shows how the DCT transform coefficients relate to a given geological model. The transform and inverse transform calculations can be carried out independently in each spatial direction and Fast Fourier Transforms (FFT) result in a relatively low computational complexity on the order of $O(P \log(P))$.

By inverting DCT coefficients at or below the truncation level $P_t$ only while setting higher order coefficients to zero, one can derive an unbiased estimate of the variance properties of a given inverse problem at a uniform spatial resolution, $R_t = \left(\dfrac{P-1}{P_t-1}\right)\Delta x$, throughout the model domain. The word "unbiased" is used here to indicate that, for any given model, the values of the DCT coefficients below the truncation level are unaffected by the choice of truncation level. It is important to note that our use of DCT differs from that in Jafarpour et al. (2009) who used this method to solve a deterministic inversion problem using bases with the largest DCT coefficients. In our work, we estimate the full range of possible models that honor the observed data at a given uniform spatial resolution.

For the DCT inversion, we consider a uniform and very densely discretized 2-D grid with a resolution that is much finer than the resulting inverse models. Instead of performing a computationally infeasible global search in the full parameter space of dimension $P \times P$, we assume that the properties of the model can be adequately described with a much lower dimensionality $P_t \times P_t$ with the remaining entries being zero. If needed, $P$ and $P_t$ can be chosen differently in each spatial direction. The definition of an appropriate size of $P_t$ is non-trivial and a simple criterion such as fitting the data to the estimated measurement errors is insufficient. Truncated and thereby disregarded DCT coefficients that should actually be non-zero, could lead to spectral leakage that affects the estimates of the retained coefficients (e.g., Trampert and Snieder, 1996; Chiao and Kuo, 2001). As a remedy to this problem, Chiao and Kuo (2001) suggest to solve for the finest level of detail possible and retrieve the model estimates corresponding to different resolutions after the inversion. In most cases, spectral leakage mainly affects the high-frequency DCT coefficients in close vicinity of the truncation limit. Consequently, a procedure based on post-inversion truncation might give more reliable (less biased) parameter estimates than if truncation is invoked at the model parameterization stage (hereafter referred to as pre-inversion truncation). The approach taken herein is thus to choose $P_t$ as large as computationally feasible and to later reconstruct lower-order models (i.e., with lower resolution) by truncating the DCT coefficients above the order considered before applying the inverse transform (referred to as post-inversion truncation in the following). This approach has the added advantage of using a single MCMC trial only, while still allowing for an analysis of the variance properties as a function of model resolution.

We determine the range of the DCT coefficients as follows. First we scale $G(1,1)$ with all other G($j,i$) that are zero to determine the bounds for which the corresponding inverse DCT models fall within $m_{min}$ and $m_{max}$. The remaining entries are scaled in a similar way so that after the inverse transform each individual coefficient has a corresponding amplitude of $\left(\dfrac{m_{\max} - m_{\min}}{2}\right)$. This ensures that all possible models can be sampled within the specified range of $m_{min}$ and $m_{max}$ at the specified resolution $R_t$. The consequence of a spectral parameterization is that we no longer invoke a uniform prior distribution of the logarithmic properties of the slowness (uniform grid discretization) but instead assume a uniform distribution of the DCT coefficients describing a transform of a logarithmic representation of slowness. This results in an Irwin-Hall distribution that resembles a Gaussian distribution at high values of $P_t$. This is not necessarily a problem as it is common practice to assume log-normal distributions of geophysical properties, but it should be kept in mind when comparing the inversion results. Another disadvantage of this parameterization is that—at least in the beginning of the MCMC inversion—many proposed DCT models will predict velocities outside the range of $m_{min}$ and $m_{max}$. Such proposals are assigned a very low log-likelihood value, and thus automatically discarded.

*Deterministic Inversions*

For comparative purposes, we calculate representative solutions based on classical smoothness-constrained iterative deterministic inversions. For this type of inversion, the objective function contains two main terms, one quantifying the data misfit and another one summarizing the difference from a prior model or pre-supposed model morphology. The inverse problem is solved iteratively by linearizing the nonlinear problem around the model obtained at the previous iteration. A trade-off parameter that defines the relative weights of the data misfit and model regularization term in the objective function is varied in a predefined manner until a model is found that explains the assumed statistics of the error model with the most parsimonious model structure possible. The inversion algorithm used herein is discussed in detail elsewhere, and we refer to Linde et al. (2006a) for the general inversion framework. Linde et al. (2008) describes how to calculate corresponding point-spread functions that describe the space-averaging and hence resolution of each estimated model parameter. Laloy et al. (2012) describe how to incorporate iteratively reweighted least-squares to minimize a perturbed $l_1$ model norm, which compared to classical least-square measures resolves sharper contrasts in the model. They also describe how to evaluate the sensitivity of each model parameter to the noise statistics of the data. For comparison with the MCMC results, we present the final inverse models, their corresponding ray-densities, and the parameter (model) uncertainties and point-spread functions at representative locations in the model.

## A Synthetic Example

As a synthetic example, we use a soil moisture model from Kowalsky et al. (2005). This model was constructed by simulating an infiltration experiment in a heterogeneous soil. The simulated soil moisture distribution is transformed into a radar velocity model using Topp's equation (Topp et al., 1980) (Fig. 1a). A synthetic data set of 900 observations is constructed for two different boreholes located 3 m apart using a transmitter-receiver geometry consisting of multiple-offset gathers between 0 and 3 m depth with sources and receiver intervals of 0.1 m. The resulting traveltimes calculated with the *time3d* algorithm (Podvin and Lecomte, 1991) are contaminated with zero mean uncorrelated Gaussian noise with a standard deviation of 0.5 ns. This constitutes a high-quality GPR data set, which is subsequently used to compare our MCMC inversion methodology against exact theory, but with noisy data. This first study serves as benchmark experiment to demonstrate the ability of the MCMC inversion procedure to back out the "known" soil moisture distribution. To understand the influence of modeling errors (inexact theory) on the inversion results, it would be necessary to calculate the response of the underlying model with a waveform modeling code, while continuing to use a ray-based approach to evaluate proposals. Such an investigation is outside the scope of the present contribution.

Before inversion, we investigate to what extent the main features of the 30 × 30 true model in Fig. 1a can be represented by an upscaled uniform grid model or a truncated DCT representation. The uniform grid (Fig. 1b) and truncated DCT (Fig. 1c) models are shown in Fig. 1 use 100 (10 × 10) model parameters. It is obvious that the DCT parameterization best represents the true radar velocity field. (Fig. 1a). Other models with sharper interfaces would favor the uniform grid parameterization. Figure 1d shows the 4 × 4 truncated DCT representation of the true model, in which the plume and the capillary fringe are clearly located, but the center of the plume appears somewhat lower than in the true model.

To assess the data misfit and define the associated stopping criterion for the deterministic inversions, we calculate the root mean square error (RMSE)

$$\text{RMSE} = \sqrt{\frac{1}{n}\sum_{i=1}^{n}\left(g_i(\mathbf{m}) - t_i\right)^2}, \qquad [10]$$

which should be close to the standard deviation of the data, namely, 0.5 ns in the examples considered herein. Figure 1e displays the deterministic least-square inversion results for isotropic smoothness constraints. The model is a rather poor representation of the true velocity model (Fig. 1a) despite the fact that the data are fitted very well (RMSE of 0.495 ns). Earlier solutions with RMSE values of about 0.53 ns (not shown herein) provided a much better representation of the plume model, highlighting the problem of which stopping criteria to use for the deterministic inversion method (outside the scope of this contribution). A better and more stable solution is offered by an inversion, in which we penalize an approximation of the $l_1$ model norm using iteratively reweighted least-squares (e.g., Farquharson et al., 2008; Laloy et al., 2012). The model (Fig. 1f) has a RMSE of 0.500 ns and the tracer plume is clearly defined, but the model is overly smooth and because of the isotropic regularization the capillary fringe zone is poorly represented.

In the MCMC inversions, we use a wide prior range of radar velocities spanning 50-170 m/μs. This causes the inverse problem to become highly nonlinear and hence challenging. The discretization for the forward model simulations is 0.10 m × 0.10 m. The 10 × 10 uniform grid discretization did not appropriately converge for this model, despite allowing twice the number of function evaluations (2 million) compared to the DCT runs. These convergence problems are likely due to abrupt changes in the properties of adjacent cells, which causes numerical instabilities in the forward simulations (Podvin and Lecomte, 1991). Indeed, numerical experiments with MCMC inversion of other geophysical data (e.g., electrical resistance tomography) involving diffusion type processes that are less sensitive to sharp cell variations demonstrate much better convergence rates with uniform grid parameterizations. Clearly, MCMC chains that have not officially converged to a limiting distribution only provide approximate information about the posterior distribution. Note that these sharp transitions are smoothed out in the DCT parameterization. Figures 2a-c show three posterior realizations based on the 10 × 10 uniform grid discretization. These different models are extremely variable and it is not particularly easy to relate these realizations to the true underlying model. The posterior mean obtained by averaging the last 1 million MCMC samples outlines a slower zone in the middle of the model, which is difficult to distinguish from the capillary fringe zone. Figures 2e-g depict three posterior realizations of the 10 × 10 DCT inversion. These models display considerable small-scale variability that is not present in the true model, with large variations between the different posterior realizations. However, all realizations correctly pinpoint the presence of low velocity zones at the actual location of the tracer plume and the capillary fringe, whereas the remaining areas mainly exhibit high velocities. The correspondence between the inversion results and true model is even better highlighted in Fig. 2h that depicts the posterior mean of the MCMC samples. The main features of the true model are clearly recovered when drawing realizations using post-inversion truncation of the results of order 4. The three example realizations in Figs. 2i-k correctly identify the location of the plume and the capillary fringe with posterior mean (Fig. 2l) similar to the optimal truncated model at this order (Fig. 1d).

The RMSE for the 10 × 10 uniform grid (Fig. 3a) and 10 × 10 DCT (Fig. 3b) MCMC inversions illustrate that both models fit the data well. The uniform grid parameterization leads to models with a slightly higher RMSE than their DCT counterparts, some of which fit the data better than the measurement error (0.5 ns). This overfitting is not surprising, but a direct consequence of the classical Gaussian likelihood function used in our MCMC inversions. This likelihood function is purposely designed to minimize the weighted sum of squared errors (see third term in Eq. [4]) without consideration of any other properties (smoothness) of the final inverted velocity field and/or soil moisture distribution.

To appraise our inversion results, we need to define an appropriate benchmark model to compare our results against. This model should not be the true model, but instead be the best possible representation of the model for a chosen basis function and model dimension. The solid red line in Fig. 4a plots the correlation coefficient of the true model and the reduced-order DCT representation of this model. The correlation coefficient rapidly increases for the first four orders, and then asymptotically approaches a value of 1. Figure 4a shows in black, for each order, the range of the correlation coefficients for the post-inversion truncated posterior model realizations of the 10 × 10 DCT models. Up to order 4, models are found with correlation coefficients that are close to optimal (the red solid line), but this correlation subsequently decreases from order 6 onwards. The blue lines represent the mean values and the range of the correlation coefficients for the pre-inversion truncations of order 1 to 9. The agreement with the post-inversion truncation results is overall high considering the presence of stochastic fluctuations between the different inversion runs. The largest differences between the pre- and post-inversion truncation models are observed for orders 2 and 3. We attribute this behavior to spectral leakage.

Classical smoothness-constrained deterministic inversions will always tend to underestimate the spatial variability of the true underlying physical property field. The absence of an explicit model regularization term in the likelihood function used herein (except for the inherent regularization associated with the order truncation), might lead to an overestimation of the spatial variability. The model structure is quantified herein through a difference operator that operates on the radar velocity model. The model structure of the truncated representations of the true model (Fig. 4b) shows a similar behavior as previously illustrated for the correlation coefficients, with a rapid increase up to order 4. Beyond this, only marginal increases occur. The posterior mean model structure derived from the MCMC samples agrees well with the true model for orders lower than 5. After this, the MCMC derived model starts to deviate considerably from the true model structure. Note that for orders 8 and higher, we did not find a single realization with equal or less model structure than the true model. This is a direct consequence of the likelihood function used in eq. [4].

The functional relationship between the RMSE and corresponding truncation order differs substantially between the pre- and post-inversion truncation results (Fig. 4c). In agreement with Laloy et al. (2012), the pre-inversion models fit the data much better than their corresponding truncations of the true model. This is an important finding, and demonstrates that these respective models do not represent samples from the actual posterior distribution. The relationship between the RMSE and truncation order is quite different for the post-inversion truncations, and demonstrates a much poorer fit for lower order posterior realizations. Yet, the best fitting models of the post-inversion truncations are in close agreement with those found for the true truncated models. Note that the use of higher order models with more parameters significantly increases the number of function evaluations to explore the posterior target distribution. Figure 4d shows how the computational demand varies with DCT order.

The results in Fig. 4 are indeed revealing. The post-inversion truncation clearly avoids overfitting of the data that prevails in pre-inversion truncation (Fig. 4c). This provides strong support for the use of a spectral representation of the model with upscaled realizations that are less amenable to overfitting. The presence of overfitting in the pre-inversion truncation results is most evident for orders 2 and 3. The correlation coefficients of these orders are noticeably lower than their counterparts derived from post-inversion truncation (see Fig. 4a). As expected, model variability increases with resolution, and we thus expect that for increasingly higher orders, the likelihood function used herein will result in models that more and more overestimate the actual heterogeneity observed in the field. To overcome this, a regularization term should be added to the likelihood function. Yet, in the absence of detailed prior information about the heterogeneity of the actual field site, it remains particularly difficult, if not impossible to decide which level of model variability is most realistic. One alternative is to turn our attention to the results of the upscaled lower-dimensional models. We postulate that a good choice of the model truncation order is one where the RMSE values of the pre-inversion truncation coincide with those found for the (best fitting) post-truncation inversion. For our synthetic case-study, this is the case for orders 4-6 (Fig. 4c), with MCMC results that compare well with the true truncated models depicted in Fig. 4d. The applicability of this finding to other geometries and geophysical data types warrants additional investigation.

To further illustrate how the variance and resolution properties of the final model vary with order, we transformed the post-inversion truncation results into soil moisture values. If we assume the widely used CRIM equation (Birchak et al., 1974) to be valid, we can directly relate the changes in GPR slowness signal to variations in soil moisture (e.g., Laloy et al., 2012) using

$$\Delta\theta = \frac{\Delta s \cdot c}{\sqrt{\kappa_w} - \sqrt{\kappa_a}}, \qquad [11]$$

where $\Delta\theta$ ($\Delta s$) indicates variations in soil moisture (slowness), $c$ denotes the speed of light in a vacuum, and $\kappa_w$ ($\kappa_a$) signifies the relative permittivity of water (air). This CRIM scaling relation uses parameters that are well known, and their uncertainty (not treated herein) should have a minor effect on the soil moisture estimates. Using Eq. [11] we can calculate the standard deviation of the soil moisture, std($\theta_{i,j}$) at location *i, j* in our 2-D modeling domain as follows

$$std\left(\theta_{i,j}\right) = \frac{std\left(s_{i,j}\right) \cdot c}{\sqrt{\kappa_w} - \sqrt{\kappa_a}}, \qquad [12]$$

where std($s_{i,j}$) denotes the corresponding standard deviation of the slowness. This mapping can easily be adapted to include uncertainty in the petrophysical relationship as well. The standard deviation of soil moisture distribution inferred from the order 4 model is shown in Fig. 5a, whereas Fig. 5b-c depicts, for two different positions in the vadose zone system, the (linear) correlation coefficients of the soil moisture with adjacent cells. If we assume that the petrophysical relationship is adequate, the soil moisture errors are relatively small <5 %, (i.e., the error in soil moisture is < 0.05 cm$^3$/cm$^3$). The largest errors are found in close vicinity of the water plume and the capillary fringe, which is consistent with the low ray coverage in these low-velocity regions. The presence of large zones with positive correlation around the cells of interest is due to the smoothness of the DCT model parameterization. The surrounding zones with negative soil moisture correlations demonstrate the presence of trade-off in the MCMC inversion. For nearby zones with very similar GPR rays, high and low velocities alternate. For higher orders, such as 6 (Figs. 5d-f), 8 (Figs. 5g-i) and 10 (Figs. 5j-l), we find soil moisture errors up to 12%, with spatial correlation images that demonstrate a rather complex trade-off in the estimated soil moisture. Note that we can easily adapt our parameter vector to include uncertainty in the estimated petrophysical model and parameters as well. Clearly, any attempts to use these results in a hydrologic context should consider models sampled at the same scale of resolution and must consider positive and negative cell-to-cell soil moisture correlations with surrounding regions. The higher the resolution of the inversion model the larger the number of soil moisture values that are being estimated, but at the expense of a larger uncertainty. This is an intuitive result as the information content of the data is finite.

We now contrast these results with those obtained by a deterministic appraisal of the inversion results in Fig. 1f. Figure 6a displays the ray density for this specific model, which illustrates a three order of magnitude variation in ray-density, with highest ray density in the high velocity region in-between the slow regions made up of the tracer plume and the capillary fringe. To assess how data errors affect the deterministic inversion results, we repeatedly (500 times) perform one additional iteration step starting from the final model, but each time with a different vector of residuals drawn randomly from the "true" error distribution. Similarly to Laloy et al. (2012), we found that the resulting soil moisture errors are very low, and range between 0.2 - 0.7% for the vast majority of the model region (Fig. 6c). These errors are unrealistically small, but easily understood as the soil moisture intervals derived this way essentially correspond to classical linear confidence intervals, and the result of a relatively poor (low) resolution of the model estimates, and use of smoothness constraint in the deterministic inversion that prohibits the trade-off fluctuations found for the MCMC inversions. Figs. 6e and 6g present point spread functions for the same two locations as previously used for the MCMC inversion. These two plots illustrate how the averaging region varies in space.

**A Field Example: South Oyster Bacterial Transport Site**

The next example concerns a field study in a saturated aquifer with variations in radar velocity that are much smaller than those previously observed for the synthetic water tracer example. We use the same borehole transect (S14-M3) as Linde et al. (2008) from the South Oyster Bacterial Transport Site (Hubbard et al., 2001; Scheibe et al., 2011). This site is referred to as Oyster in what follows next. The data were acquired using a PulseEKKO 100 system with 100-MHz nominal-frequency antennae. A total of 3,248 traveltimes were picked using a transmitter and receiver space step of 0.125 m in each borehole. The deterministic inversion of these high-quality data to a RMSE of 0.5 ns resulted in velocity models with a strong correlation (0.72) with the logarithm of permeability inferred from electromagnetic flowmeter data acquired in M2 (Linde et al., 2008).

Deterministic inversion results of the Oyster data are shown in Fig. 7a for isotropic smoothness constraints, and for anisotropic smoothness constraints (anisotropy factor of 5) using an $l_2$ (Fig. 7b) and $l_1$ (Fig. 7c) model norm. The most realistic models are found when using an anisotropic regularization. Flowmeter data from the experimental site suggests a horizontal anisotropy ratio of about 5 (Hubbard et al., 2001), which cannot be retrieved with isotropic smoothness constraints. The RMSE of these models are 0.505, 0.502, and 0.496 ns, respectively. These values do not constitute the lowest possible RMSE values (i.e., lower RMSE values can be achieved), but correspond to models that are consistent with the expected noise level.

The MCMC inversion region was defined as a 7.4 m × 7.4 m model domain that was parameterized as a 10 × 10 inversion grid (roughly one wavelength), or alternatively using the first 10 × 10 DCT coefficients, while the forward simulations were carried out on a 40 × 40 grid. We deliberately limited the inversion to 10 × 10 = 100 parameters so that the required MCMC computing time remains reasonable. The MT-DREAM$_{(ZS)}$ code will also converge when using 400 parameters (half a wavelength), but the time it takes to converge to a limiting distribution significantly increases. The prior velocity range was chosen to be 50-70 m/μs, which is considerably smaller than the range used for the synthetic benchmark study described previously, yet significantly larger than the range obtained from deterministic inversions (Linde et al., 2008). We first carried out MCMC inversions in which the parameters of the uniform grid and DCT models were allowed to vary freely within their prior defined ranges, hereafter referred to as unconstrained inversion. We subsequently performed additional MCMC inversions in which models that did not display at least 5 times more roughness (i.e., the sum of squares of gradients in the model) in the vertical direction compared to the horizontal direction were penalized. In the remainder of this paper, these runs are referred to as anisotropy-constrained inversion.

The data misfits of the unconstrained MCMC inversion based on a 10 × 10 uniform grid parameterization resulted in mean RMSE values of about 0.579 ns. On the contrary, the 10 × 10 DCT inversion provided RMSE values of about 0.475 ns (Fig. 3d). Indeed, the DCT parametrization receives the best performance for this data set, as both methods use the exact same number of parameters, and hence effective resolution. The RMSE of the corresponding anisotropic uniform grid model is about 0.609 ns (Fig. 3e), whereas their DCT counterparts exhibit smaller RMSE values of approximately 0.496 ns (Fig. 3f). This finding explains why we cannot find a single model with lateral anisotropy in the unconstrained inversion results. Indeed, a model with a RMSE of 0.496 ns (A) is too far from the best model of 0.475 ns (B). The jump (Metropolis) probability, $p(B \rightarrow A)$ is less than $10^{-57}$, a direct consequence of the high number of GPR observations used in the log-likelihood function.

The MCMC inversions based on uniform grid models were stopped prematurely after 600,000 function evaluations as the RMSE values stagnated and did not show further improvements. Even if formal convergence might have been declared at a later stage, the RMSE values are unrealistically high and cannot result in credible models. Again, we mainly attribute this poor convergence of the uniform grid representation to the inability of the forward model to efficiently handle sharp interfaces. Formal convergence was declared after about 603,000 model evaluations for the unconstrained DCT inversion and after approximately 270,000 successive model evaluations for the anisotropy-constrained inversion.

Posterior realizations of the unconstrained MCMC inversions (Figs. 7d-f for the 10 × 10 uniform grid discretization; Figs. 7g-i for the 10 × 10 DCT) do not adequately represent the expected horizontal layering, but bear some similarity with the model obtained using isotropic smoothness constraints (Fig. 7a). The corresponding anisotropy-constrained uniform grid models display similar features as the anisotropy-constrained deterministic inversions (Figs. 7b-c), but appear very granular with considerably higher mean RMSE values of about 0.609 ns. The anisotropy-constrained DCT MCMC inversions provide well-constrained models (Figs. 7m-o) that are visually very similar to those obtained by the anisotropic smoothness-constrained deterministic inversion (Figs. 7b-c). One subtle difference is that the anisotropy-constrained DCT results derived from MCMC simulation provide an improved geological continuity of high- and low velocity regions.

The anisotropy-constrained DCT inversion results were also analyzed in terms of their estimates of porosity uncertainty and spatial correlation. This is done in a similar fashion as for the synthetic case study. We find that the estimated errors are very low (Fig. 8a), in fact lower than 0.2 % in most of the central region of the model domain. This finding fundamentally differs from the results of the synthetic soil moisture plume, and is explained by the increase in data (3,248 vs. 900 observations), the presence of a smaller velocity range that decreases non-linear effects dramatically, and the importance of the anisotropy constraints. Indeed, the models without anisotropy constraints demonstrate a much larger variability (Figs. 7g-i) and fit the data better (c.f., Figs. 3d and 3f).

The deterministic inversion results (Fig. 7c) indicate the presence of a more evenly distributed ray-coverage (Fig. 6b) compared to the synthetic water tracer example (Fig. 6a). This is a direct consequence of the smaller velocity range in the Oyster data. The estimated model uncertainties (Fig. 6d) are even smaller than those for the synthetic case (Fig. 6c) with most of the model regions having soil moisture errors smaller than 0.07 %. These errors are only about two times smaller than those of the MCMC results. This demonstrates that the differences between deterministic and stochastic inversion diminishes with decreasing nonlinearity of the forward model. Two point-spread functions (Figs. 6f and 6h) illustrate that despite the small velocity contrasts, the lower high velocity zone in Fig. 7c is considerably better resolved than the upper low velocity zone.

**Discussion**

We have presented a Bayesian inversion framework that estimates the two (or three) dimensional posterior soil moisture distribution from GPR travel time observations. The methodology presented herein is based on a formal likelihood function, and should provide vadose zone hydrology with an improved and statistically sound procedure to estimate spatially distributed soil moisture values, and their underlying uncertainty and correlation at different spatial resolutions (see Figs. 5 and 8). These estimates could then serve as data for hydrologic modeling and parameter estimation problems, such as those treated by, among others, Cassiani et al. (1998), Chen et al. (2001), and Farmani et al. (2008). To determine the main features of the vadose zone system, it might be most productive to start with a low variance estimate at a low spatial resolution. High-resolution and larger variance models can then be used at a subsequent stage to help resolve small-scale details. The petrophysical relationship and its associated uncertainty should be expressed at the same scale as the resolution of the inversion model, requiring an appropriate upscaling (outside the scope of the present paper). The methodology presented herein is complementary to fully-coupled hydrogeophysical inversion (e.g., Kowalsky et al., 2005) as the posterior soil moisture distribution derived with our approach will help with (a) the identification and construction of an adequate conceptual model, (b) model parameterization, and (c) determination of the initial conditions. The methodology should also shed more light on the importance and treatment of model structural deficiencies. More research is warranted in this direction.

The applications presented herein are based on the following three main assumptions that can be relaxed in future studies: (1) a two-dimensional model adequately describes the GPR observations, (2) data errors are spatially and temporally uncorrelated, and (3) an asymptotic solution is sufficient to simulate EM wave propagation. These assumptions might not be completely realistic. For instance, ray-bending seems likely to take place outside the plane defined by the two boreholes, GPR measurement errors might be spatially correlated, and the asymptotic solution to simulate the propagation of EM waves through the vadose zone system under consideration might not always be sufficiently accurate. A three-dimensional and full waveform modeling procedure would resolve two of the three main limitations of the presented procedure, but at the expense of a significant increase in computational costs. We posit that this approach will significantly reduce the model variance at high spatial resolutions and decrease the risk of biased parameter estimates due to model errors. Another way to decrease model variance is to add other geophysical data types to the inverse analysis. For example, we suspect that additional conditioning to GPR reflection data could be effective.

We consistently found that convergence was superior for DCT models compared to uniform grid parameterizations, which in fact, did not formally converge within the computational budget for the models and modeling code considered herein. Consequently, we might have reported RMSE values that are somewhat larger than those of the true posterior models. The convergence problems of the uniform grid models is caused by abrupt variations in the properties of neighboring cells that are introduced during the inversion, and pose significant challenges for the numerical solver of the forward problem. Other tests (not shown herein) demonstrate that the results for the uniform grid discretization are improved if a moving average filter is used to smoothen neighboring cells in the grid before each successive forward simulation. This made us favor the DCT as this model parameterization ensures continuity.

The results presented herein allow us to outline a number of key questions that should be addressed in future studies, including (1) how do we best define the prior ranges of the transformed model parameterizations? (2) how do we derive upscaled petrophysical relationships at a given scale and what are the associated spatial correlation structures of these relationships? (3) what is the effect of incomplete geometrical information and incorrect physics in the forward model on the simulation results and data misfit? (4) how do we diagnose, detect and resolve model errors originating from model truncation? (5) how do we define an appropriate spatial resolution of our inverse model which is consistent with the information content of the GPR observations?

Finally, for high-dimensional applications it is particularly important to develop more advanced likelihood functions that avoid excessively granular and variable models, and overfitting of the data. The use of informative priors might help in this regard (e.g., Cordua et al., 2012), yet information about the expected model variability remains necessary to derive credible modeling results. This can be problematic for many field applications when such prior information is not readily available.

## Conclusions

Computer-intensive MCMC inversions must often seek model parameterizations that can explain most of the expected complexity and heterogeneity of the physical property of interest with the least number of parameters. A parameterization that is too simple leads to an unacceptable bias or overly simplified model structure (e.g., a layered model with uniform layer properties). On the contrary, a model parameterization that is too complex might be computationally infeasible, and demonstrate too much variability with posterior solutions that poorly represent the true soil moisture distribution.

We have investigated the usefulness and applicability of the presented Bayesian inversion methodology using two differing and opposing model parameterization schemes, one involving a classical uniform grid discretization and the other based on DCT. The DCT parameterization exhibits superior results for the smoothly varying property field considered herein, not only in terms of an improved data fit with visually superior models and more realistic estimates of model (soil moisture) uncertainty over a wide range of spatial resolutions, but also in terms of MCMC convergence speed. The Bayesian modeling uncertainties were found to be much larger than those obtained from a classical deterministic inversion. The post-inversion truncation strategy, in which the posterior models are truncated at different levels to investigate the trade-off between resolution and variance, was deemed successful and helped to avoid data overfitting. A truncation level defined where the lowest data misfits of the pre- and post-inversion truncation results coincide appears to represent an adequate level of model complexity. For the field example, it was necessary to add additional constraints in the inversion to enforce the lateral anisotropy observed in borehole data. No such layering was observed in the unconstrained MCMC inversion results.

Future work would be most productive if focused on providing better guidelines and methods on how to build more realistic models in transform domains. Borehole data and/or training images will help to restrict the feasible parameter space, and improve the resemblance of the posterior inversion results with the actual field situation. Further improvements to the likelihood function are warranted to assure that truncated representations of the true model are consistent with samples from the posterior distribution derived from pre-inversion truncation MCMC simulation. Other more intermediate model parameterization strategies containing both spatial and spectral localization, such as wavelets, require further investigation.

*Acknowledgements*

We are grateful to John Peterson and Susan Hubbard who provided the crosshole GPR data from the South Oyster Bacterial Transport Site and Michael Kowalsky who shared the synthetic soil moisture model. We thank Eric Laloy for many stimulating discussions. Comments from the AE Sébastien Lambot and two anonymous reviewers are greatly appreciated and helped to improve the manuscript. A MATLAB version of the MT-DREAM$_{(ZS)}$ code used herein can be obtained from the 2$^{nd}$ author upon request (jasper@uci.edu). This code includes a simplified version of the synthetic example considered herein.

## References

Ahmed, A.T., T. Natarjan, and K.R. Rao. 1974. Discrete cosine transform. IEEE T. Bio.-Med. Eng. C23:90-93.

Alumbaugh, D., P.Y. Chang, L. Paprocki, J.R. Brainard, R.J. Glass, and C.A. Rautman. 2002. Estimating moisture contents in the vadose zone using cross-borehole ground penetrating radar: A study of accuracy and repeatability. Water Resour. Res. 38:1309. doi:10.1029/2001WR000754

Alumbaugh, D.L., and G.A. Newman. 2000. Image appraisal for 2-D and 3-D electromagnetic inversion. Geophysics 65:1455-1467. doi:10.1190/1.1444834

Bikowski, J., J. A. Huisman, J. A. Vrugt, H. Vereecken, and J. van der Kruk. 2012. Integrated analysis of waveguide dispersed GPR pulses using deterministic and Bayesian inversion methods. Near Surface Geophysics. doi: 10.3997/1873-0604.2012041

Binley, A., P. Winship, R. Middleton, M. Pokar, and J. West. 2001. High-resolution characterization of vadose zone dynamics using cross-borehole radar. Water Resour. Res. 37:2639-2652. doi:10.1029/2000WR000089

Birchak, J.R., C.G. Gardner, J.E. Hipp, and J.M. Victor. 1974. High dielectric constant microwave probes for sensing soil moisture. Proc. IEEE. 62:9398. doi:10.1109/PROC.1974.93–88

Cassiani, G., G. Böhm, A. Vesnaver, R. Nicolich. 1998. A geostatistical framework for incorporating seismic tomography auxiliary data into hydraulic conductivity. J. Hydrol. 206:58-74. doi:10.1016/S0022-1694(98)00084-5

Chen, J.S., S. Hubbard, and Y. Rubin. 2001. Estimating the hydraulic conductivity at the South Oyster Site from geophysical tomographic data using Bayesian techniques based on the normal linear regression model. Water Resour. Res. 37:1603-1613. doi:10.1029/2000WR900392

Chiao, L.-Y., and B.-Y. Kuo. 2001. Multiscale seismic tomography. Geophys. J. Int. 145:517-527. doi:10.1046/j.0956-540x.2001.01403.x

Constable, S.C., R.L. Parker, and C.G. Constable. 1987. Occam's inversion: A practical algorithm for generating smooth models from electromagnetic sounding data. Geophysics 52:289-300. doi:10.1190/1.1442303

Cordua, K.S., T.M. Hansen, and K. Mosegaard. 2012. Monte Carlo full-waveform inversion of crosshole GPR data using multiple-point geostatistical a priori information. Geophysics 77:19-31. doi:10.1190/GEO2011-0170.1

Dafflon, B., J. Irving, and K. Holliger. 2009. Simulated-annealing-based conditional simulation using crosshole georadar for the local-scale characterization of heterogeneous aquifers. J. Appl. Geophys. 68:60-70. doi:10.1016/j.jappgeo.2008.09.010

Day-Lewis F.D., and J.W. Lane Jr. 2004. Assessing the resolution-dependent utility of tomograms for geostatistics. Geophys. Res. Lett. 31:L07503. doi:10.1029/2004GL019617

Day-Lewis F.D., K. Singha, and A.M. Binley. 2005. Applying petrophysical models to radar travel time and electrical resistivity tomograms: Resolution-dependent limitations. J. Geophys. Res. 110:B08206. doi:10.1029/2004JB003569

Ernst, J.R., A.G. Green, H. Maurer, and K. Holliger. 2007. Application of a new 2D time-domain full-waveform inversion scheme to crosshole radar data. Geophysics 72:J53-J64. doi:10.1190/1.2761848

Ernst, J.R., K. Holliger, H. Maurer, and A.G. Green. 2006. Realistic FDTD modelling of borehole georadar antenna radiation: methodolgy and application. Near Surf. Geophys. 4:19-30.

Eppstein, M.J., and D.E. Dougherty. 1998. Efficient three-dimensional data inversion: Soil characterization and moisture monitoring from cross-well ground-penetrating radar at a Vermont test site. Water Resour. Res. 34:1889-1900. doi:10.1029/98WR00776

Farmani M.B., N.-O. Kitterød, and H. Keers. 2008. Inverse modeling of unsaturated flow parameters using dynamic geological structure conditioned by GPR tomography. Water Resour. Res. 44:W08401. doi:10.1029/2007WR006251

Farquharson, C.G. 2008. Constructing piecewise-constant models in multidimensional minimum-structure inversions. Geophysics 73:K1-K9. doi:10.1190/1.2816650

Gelman, A.G., and D.B. Rubin. 1992. Inference from iterative simulation using multiple sequences. Stat. Sci. 7:457-472

Gloaguen E, D. Marcotte, B. Giroux, and H. Perroud. 2007. Stochastic borehole radar velocity and attenuation tomographies using cokriging and cosimulation. J. Appl. Geophys. 62:141-157. doi:10.1016/j.jappgeo.2005.01.001

Hansen, T.M., M.C. Looms, and L. Nielsen. 2008. Inferring the subsurface structural covariance model using cross-borehole ground penetrating radar tomography. Vadose Zone J. 7:249-262. doi:10.2136/vzj2007.0087

Hansen, T.M., A.G. Journel, A. Tarantola, and K. Mosegaard. 2006. Linear inverse Gaussian theory and geostatistics. Geophysics 71:R101-R111. doi:10.1190/1.2345195

Hinnell, A.C., T.P.A. Ferré, J.A. Vrugt, J.A. Huisman, S. Moysey, J. Rings, and M.B. Kowalsky. 2010. Improved extraction of hydrologic information from geophysical data through coupled hydrogeophysical inversion. Water Resour. Res. 46:W00D40. doi:10.1029/2008WR007060

Hubbard, S.S., Y. Rubin, and E. Majer. 1999. Spatial correlation structure estimation using geophysical and hydrogeological data. Water Resour. Res. 35:1809-1825. doi:10.1029/1999WR900040

Hubbard, S.S., J. Chen, J. Peterson, E.L. Mayer, K.H. Williams, D.J. Swift, B. Mailloux, and Y. Rubin. 2001. Hydrogeological characterization of the South Oyster Bacterial Transport Site using geophysical data. Water Resour. Res. 37:2431-2456

Jackson, D. 1976. Most squares inversion. J. Geophys. Res. 81:1027-1030.

Jafarpour, B., Goyal V.K., W.T. Freeman, and D.B. McLaughlin. 2009. Transform-domain sparsity regularization for inverse problems in geosciences. Geophysics 74:R69-R83. doi:10.1190/1.3157250

Jafarpour, B., and D.B. McLaughlin. 2008. History-matching with an ensemble Kalman filter and discrete cosine parameterization. Comp. Geosci. 12:227-244. doi:10.1007/s10596-008-9080-3

Kalscheuer, T., and L.B. Pedersen. 2007. A non-linear truncated SVD variance and resolution analysis of two-dimensional magnetotelluric models. Geophys. J. Int. 169:435-447. doi:10.1111/j.1365-246X.2006.03320.x

Klotzsche, A., J. van der Kruk, G.A., Meles, J. Doetsch, H. Maurer, and N. Linde. 2010. Full-waveform inversion of cross-hole ground-penetrating radar data to characterize a gravel aquifer close to the thur River, Switzerland. Near Surf. Geophys. 8:635-649. doi:10.3997/1873-0604.2010054

Kowalsky M.B., S. Finsterle, S. Hubbard, Y. Rubin, E. Majer, A. Ward, and G. Gee. 2005. Estimation of field-scale soil hydraulic and dielectric parameters through joint inversion of GPR and hydrological data. Water Resour. Res. 41:W11425. doi:10.1029/2005wr004237

Laloy, E., N. Linde, and J.A. Vrugt. 2012. Mass conservative three-dimensional water tracer distribution from MCMC inversion of time-lapse GPR data. Water Resour. Res. 48: W07510. doi:10.1029/2011WR011238

Laloy, E. and J.A. Vrugt. 2012. High-dimensional posterior exploration of hydrological models using multiple-try DREAM$_{(ZS)}$ and high-performance computing. Water Resour. Res. 48:W01526. doi:10.1029/2011WR010608

Lambot, S., E.C. Lambot, I. van den Bosch, B. Stockbroeckx, and M. Vanclooster. 2004. Modeling of ground-penetrating radar for accurate characterization of subsurface electric properties. IEEE T. Geosci. Remote 42: 2555-2568. doi:10.1109/TGRS.2004.834800

Linde N., A. Binley, A. Tryggvason, L.B. Pedersen, A. Revil. 2006a. Improved hydrogeophysical characterization using joint inversion of cross-hole electrical resistance and ground-penetrating radar traveltime data. Water Resour. Res. 42:W12404. doi:10.1029/2006WR005131

Linde N., S. Finsterle, S. Hubbard. 2006b. Inversion of tracer test data using tomographic constraints. Water Resour. Res. 42:W04410. doi:10.1029/2004WR003806

Linde, N., A. Tryggvason, J.E. Peterson, S.S. Hubbard. 2008. Joint inversion of crosshole radar and seismic traveltimes acquired at the South Oyster Bacterial Transport Site. Geophysics 73:G29-G37. doi:10.1190/1.2937467

Meju, M. A., and V. R. S. Hutton. 1992. Iterative most-squares inversion: application to magnetotelluric data. Geophys. J. Int. 108:758-766. doi:10.1111/j.1365-246X.1992.tb03467.x

Moysey S., K. Singha, and R. Knight. 2005. A framework for inferring field-scale rock physics relationships through numerical simulation. Geophys. Res. Lett. 32:L08304. doi:10.1029/2004GL022152

Oldenburg, D.W., and Y.G. Li. 1999. Estimating depth of investigation in dc resistivity and IP surveys. Geophysics 64:403-416. doi:10.1190/1.1444705

Podvin, P., and I. Lecomte. 1991. Finite difference computation of travel times in very contrasted velocity models: a massively parallel approach and its associated tools. Geophys. J. Int.. 105:271-281. doi:10.1111/j.1365-246X.1991.tb03461.x

Roth, K., R. Schulin, H. Flühler, and W. Attinger. 1990. Calibration of time domain reflectometry for water content measurement using a composite dielectric approach. Water Resour. Res. 26:2267-2273.

Rubin, Y., G. Mavko, and J. Harris. 1992. Mapping permeability in heterogeneous aquifers using hydrological and seismic data. Water Resour. Res. 28:1809-1816. doi:10.1029/92WR00154

Scheibe, T. D., S. S. Hubbard, T. C. Onstott, and M. F. DeFlaun. 2011. Lessons learned from Bacterial Transport Research at the South Oyster Site. Ground Water 49:745-763. doi:10.1111/j.1745-6584.2011.00831.x

Simons, F. J., Dahlen, F. A., and M. A. Wieczorek. 2006. Spatispectral concentration on a sphere. SIAM Review 48:504-536. doi:10.1137/S0036144504445765

Tarantola, A. 2005. Inverse problem theory and methods for model parameter estimation, SIAM, Philadelphia.

ter Braak, C. J. F. 2006. A Markov chain Monte Carlo version of the genetic algorithm differential evolution: Easy Bayesian computing for real parameter space. Stat. Comput. 16:239-249. doi:10.1007/s11222-006-8769-1

Topp, G. C., J. L. Davis, and A. P. Annan (1980), Electromagnetic determination of soil water content: Measurements in coaxial transmission lines. Water Resour. Res. 16:574-582.

Trampert, J. and R. Snieder. 1996. Model estimations biased by truncated expansions: Possible artifacts in seismic tomography. Science 271:1257-1260. doi:10.1126/science.271.5253.1257

Vidale, J. 1988. Finite-difference calculation of travel times. Bulletin of the Seismological Society of America 78:2062-2076.

Vrugt, J. A., C. J. F. ter Braak, M. P. Clark, J. M Hyman, B. A. Robinson. 2008. Treatment of input uncertainty in hydrologic modeling: Doing hydrology backward with Markov chain Monte Carlo simulation. Water Resour. Res. 44:W00B09. doi:10.1029/2007WR006720.

Vrugt, J.A., C.J.F. ter Braak, C. G. H. Diks, D. Higdon, B. A. Robinson, and J.M. Hyman. 2009. Accelerating Markov chain Monte Carlo simulation with self-adaptive randomized subspace sampling. Int. J. Nonlin. Sci. Num. 10:273-290.

Warren, C. and A. Giannopoulos. 2011. Creating finite-difference time-domain models of commercial ground-penetrating radar antennas using Taguchi's optimization method. Geophysics 76, G37-G47. doi: 10.1190/1.3548506.

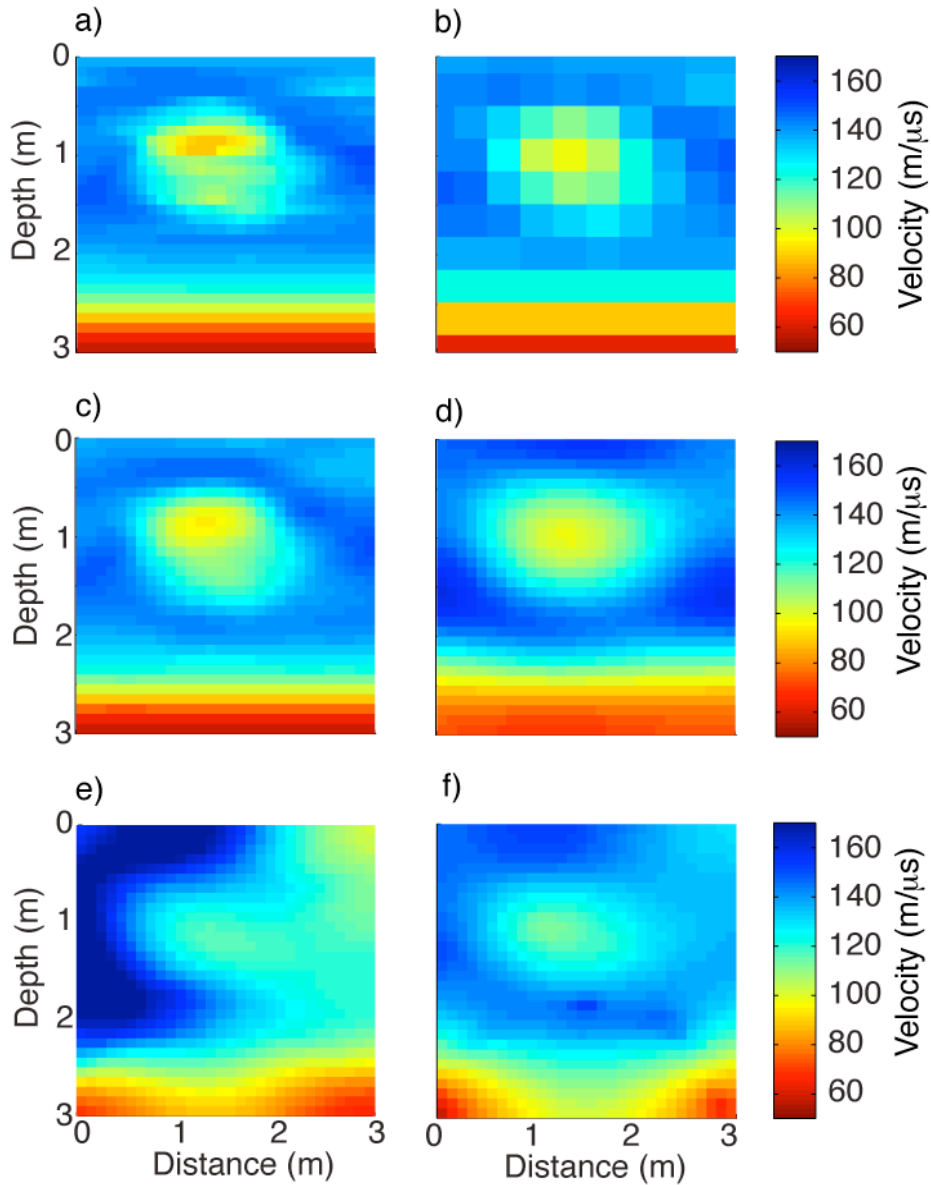

**Fig. 1.** Synthetic radar velocity model derived from the simulated soil moisture distribution of an infiltrating water tracer plume. (a) True model (Kowalsky et al., 2005), (b) upscaled uniform grid representation with 10 × 10 model parameters, (c) upscaled DCT representation with (c) 10 × 10 and (d) 4 × 4 model parameters, respectively. Velocity models obtained by deterministic inversions using isotropic smoothness constraints with (e) a $l_2$ model norm or (f) a perturbed $l_1$ model norm.

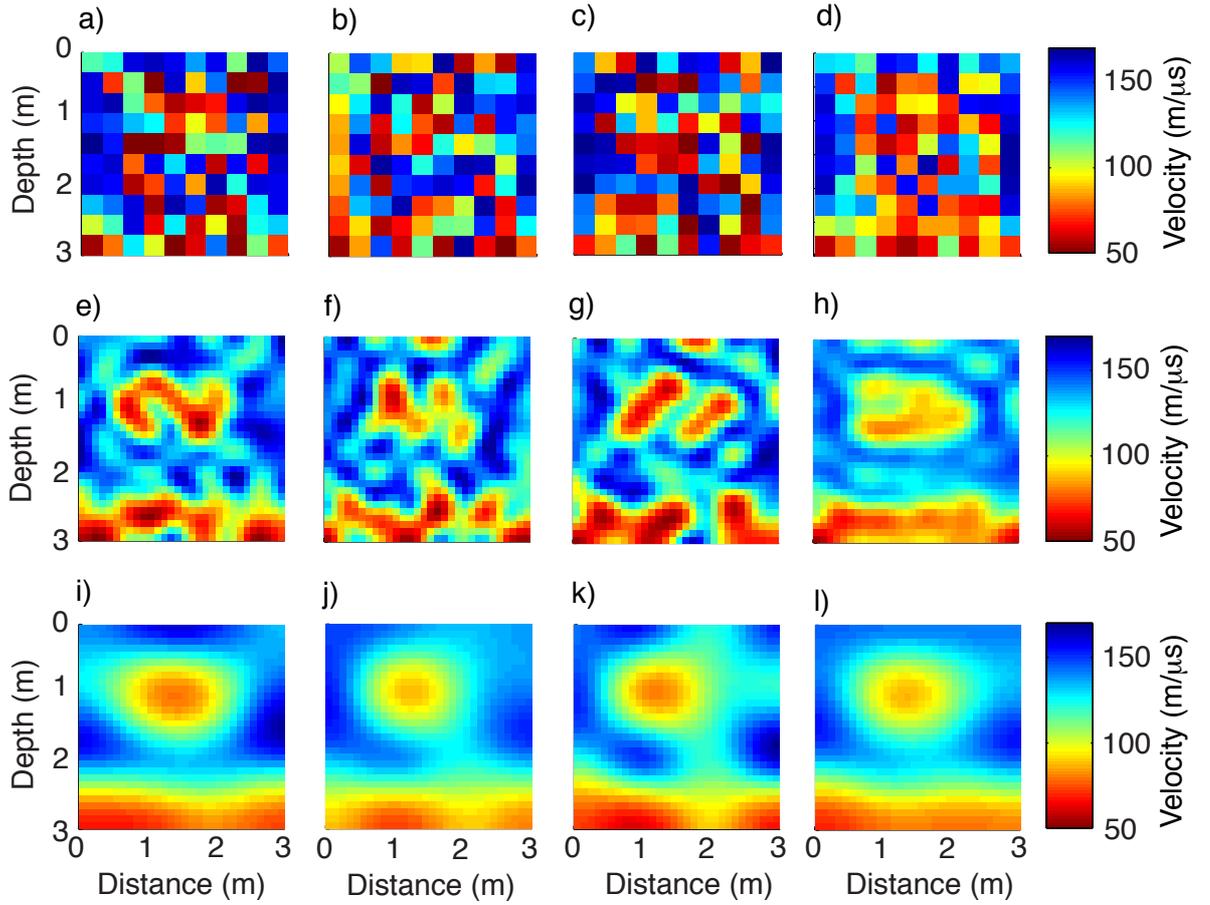

**Fig. 2.** Posterior MCMC realizations for the water tracer plume example. The first three columns (a-c), (e-g), and (i-k) depict the velocity field of three randomly selected posterior solutions, whereas the fourth panel at the right hand side (d, h, l) plots the posterior mean velocity field. The top row (a-d) corresponds to the results of the 10 × 10 uniform grid discretization, whereas the bottom two rows summarize our findings for the (e-h) 10 × 10 DCT, and (i-l) 4 × 4 DCT parameterization.

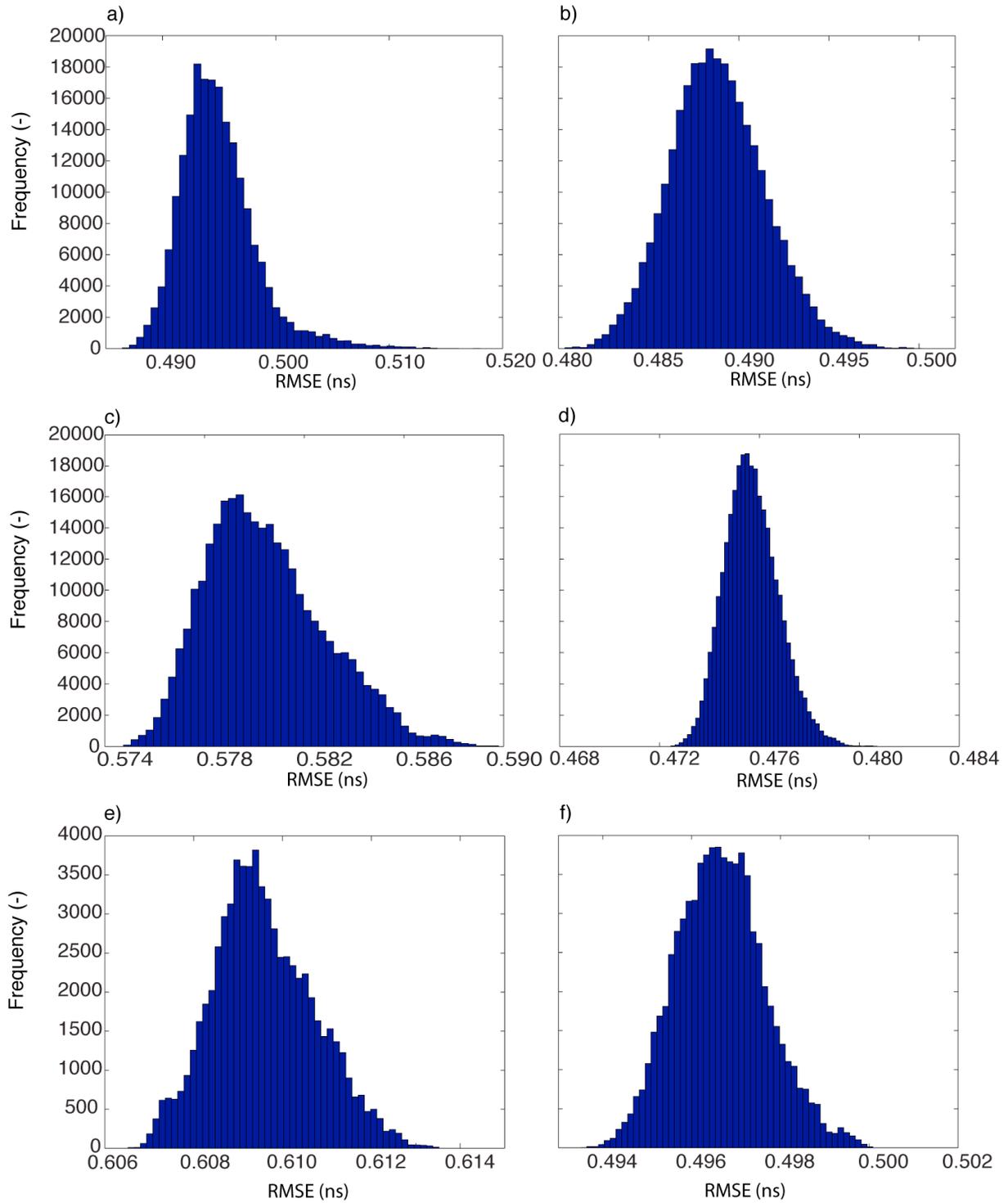

**Fig. 3.** Histograms of the sampled RMSE values of the different MCMC inversion runs: synthetic water tracer experiment with (a) 10 × 10 uniform grid discretization and (b) 10 × 10 DCT representation; (c-d) unconstrained inversion of the Oyster data with (c) 10 × 10 uniform grid discretization and (d) 10 × 10 DCT representation; (e-f) anisotropy-constrained inversion of the Oyster data with (e) 10 × 10 uniform grid discretization and (f) 10 × 10 DCT representation.

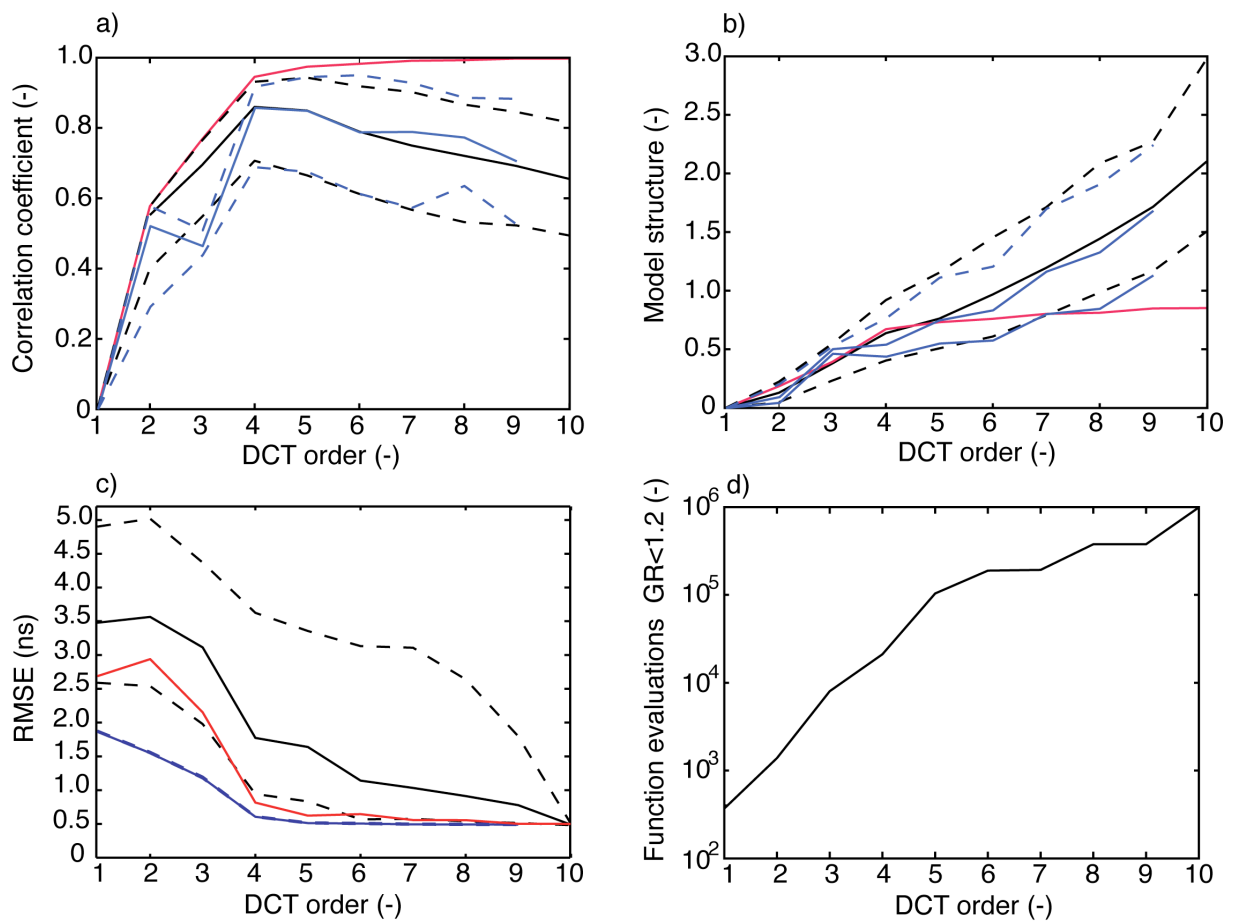

**Fig. 4.** Metrics of the MCMC inversion results for the synthetic water tracer plume considered herein as a function of DCT truncation order: (a) Correlation coefficients of proposed models and the true model, (b) the model structure, (c) RMSE of model simulations and original data, and (d) required number of MCMC function evaluations to formally converge, as a function of truncation order. Comparison statistics for truncations of the true model are shown in red, whereas blue and black lines present the results for the pre- and post-inversion truncation. The solid lines summarize the mean values, and the dashed lines represent the minimum and maximum values of the solutions.

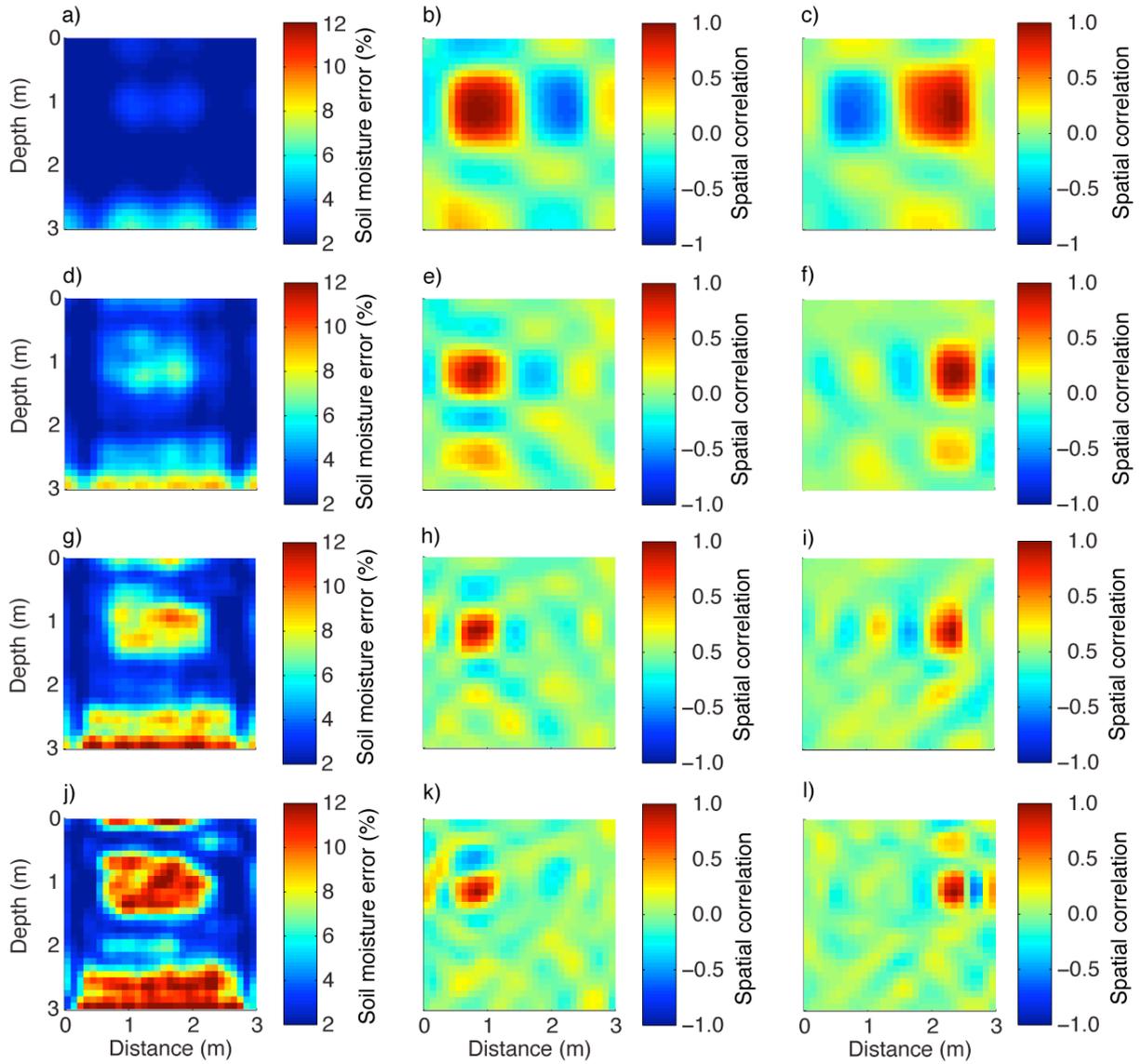

**Fig. 5.** Standard deviations (left column) of soil moisture and associated spatial correlations with respect to a cell centered at depth = 1.15 m and distance = 0.85 m (middle column) and depth = 1.15 m and distance = 2.35 m (right column) derived from the posterior samples from the MCMC inversion using DCT. Results correspond to post-inversion truncation for (a-c) order 4, (d-f) order 6, (g-i) order 8, and pre-inversion truncation for (j-l) order 10. Possible errors in the petrophysical relationship are not accounted for in this plot, but are easy to add given that their uncertainty and support scale are known (see Eq. [5]).

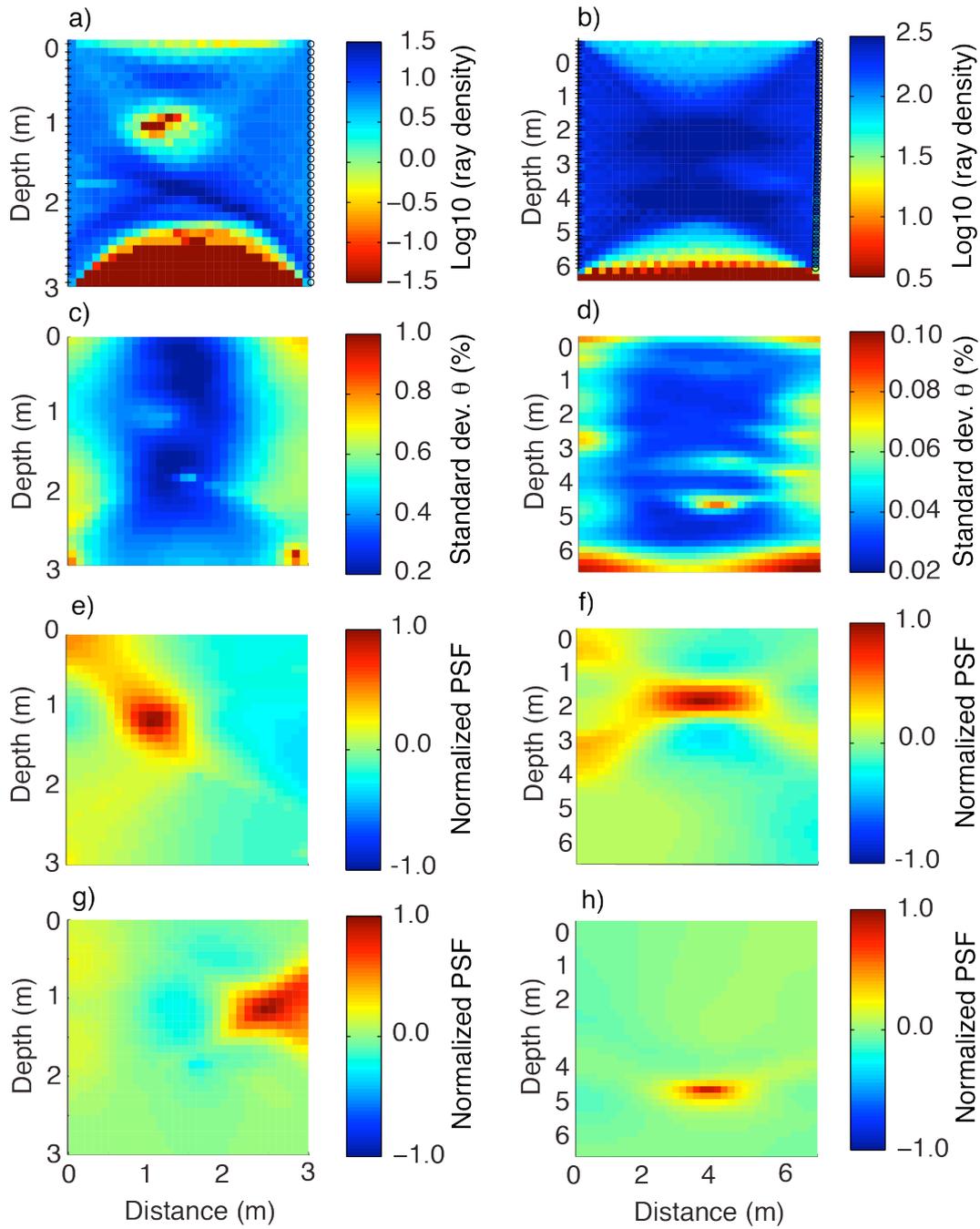

**Fig. 6.** Image appraisal of the deterministic inversion results: (a-b) Ray-density for (a) the synthetic water tracer experiment and (b) the Oyster case-study for the models shown in Figs. 1f and 5c, respectively; (c-d) Deterministic soil moisture errors for (c) the synthetic water tracer and (d) the Oyster case-study; (e-h) Example point-spread functions (PSFs) for the (e,g) synthetic and (f,h) Oyster case study: (e) cell centered at depth = 1.15 m and distance = 0.85 m and (f) depth = 1.15 m and distance = 2.35 m, respectively, and (f) cell centered at depth = 1.76 m and distance = 3.5 m and (g) depth = 4.76 m and distance = 3.5 m, respectively. The values of the PSFs are normalized to the value of the model cells of interest.

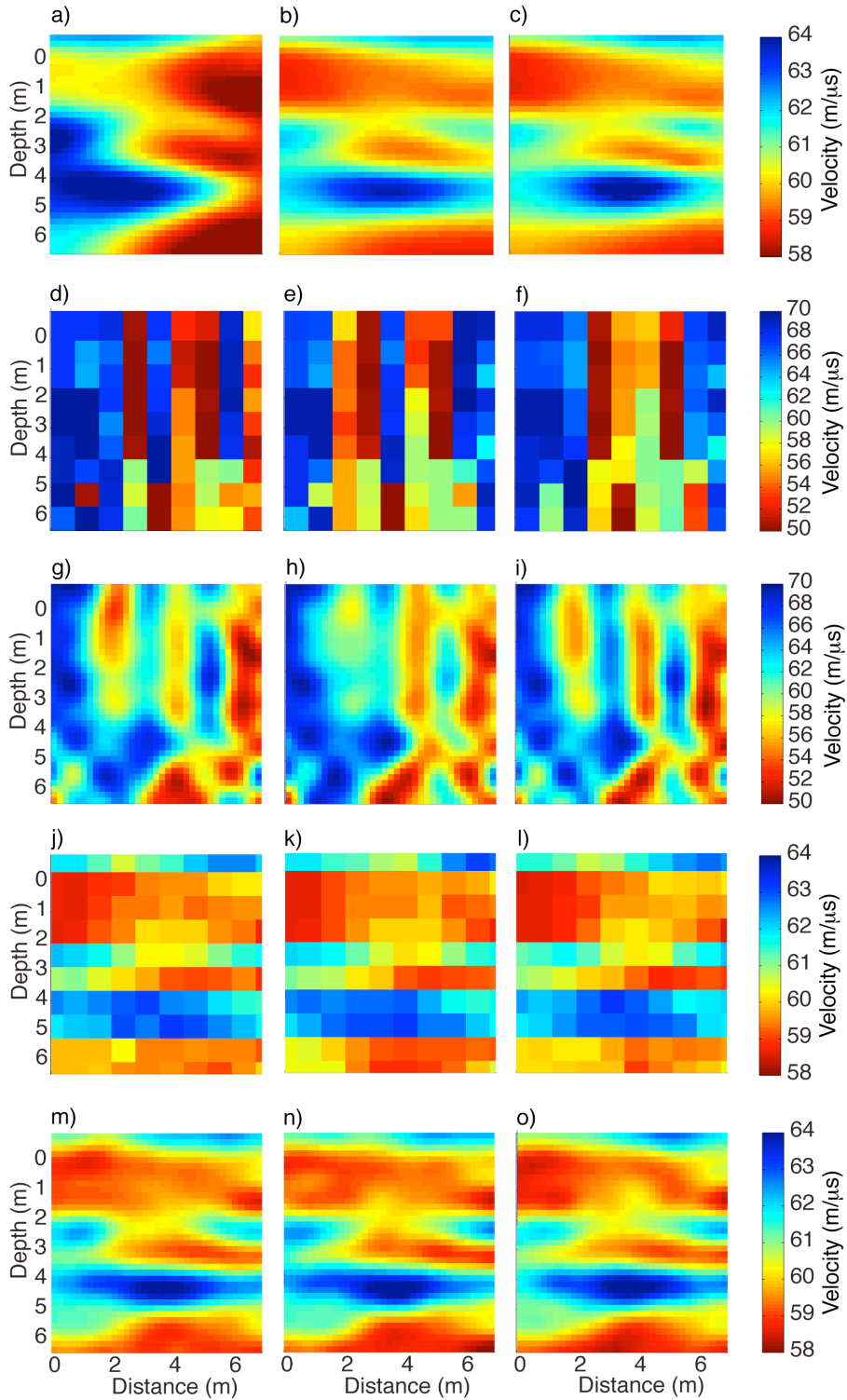

**Fig. 7.** Example models obtained by inversion of the Oyster data set: deterministic least-squares inversion with (a) isotropic smoothness-constraints, and (b-c) anisotropic smoothness-constraints (5 times) using (b) a $l_2$ model norm or (c) a perturbed $l_1$ model norm; unconstrained 10 × 10 MCMC inversion with (d-f) uniform grid discretization and (g-i) DCT parameterization; anisotropy-constrained 10 × 10 MCMC inversion with (j-l) uniform grid discretization and (m-o) DCT parameterization.

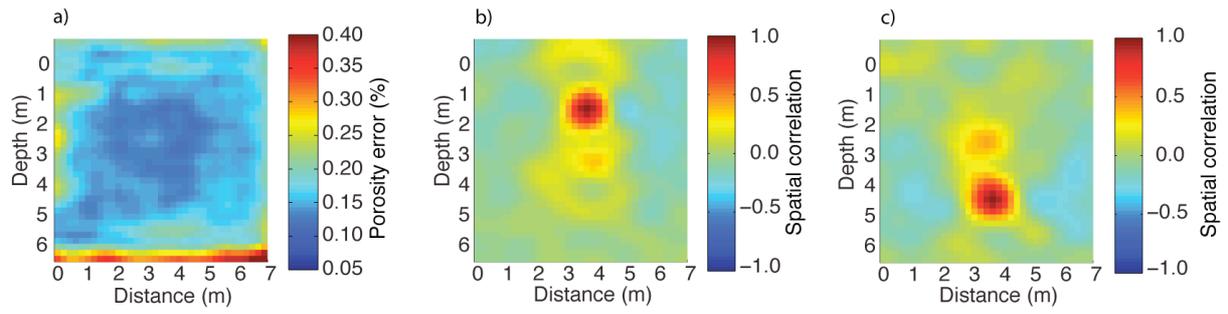

**Fig. 8.** MCMC derived standard deviations for the Oyster case study using the DCT: (a) porosity; (b-c) spatial correlations of the porosity error estimates are shown with respect to (b) a cell centered at depth = 1.76 m and distance = 3.5 m and (c) depth = 4.76 m and distance = 3.5 m. Errors in the petrophysical relationship are not accounted for in this plot, but are easy to add given that their uncertainty and support scale are known (see Eq. [5]).